\documentclass[11pt,a4paper]{article} 
\pdfoutput=1
\usepackage{jheppub}

\usepackage{amsmath, amssymb}
\usepackage{mathpazo}
\usepackage{mathrsfs}
\usepackage{array,arydshln}

\font\sm=cmss10 at 4pt

\usepackage{graphicx,epsfig}
\usepackage{epic}
\usepackage{color}
\usepackage{youngtab}

\newcommand{\be}{\begin{equation}}
\newcommand{\ee}{\end{equation}}
\newcommand{\ba}{\begin{eqnarray}}
\newcommand{\ea}{\end{eqnarray}}

\newcommand{\mc}{\mathcal }

\newcommand{\wt}{\widetilde}

\newcommand{\mk}{\mathfrak}

\newcommand{\hs}{\mbox{hs}[\lambda]}
\newcommand{\sref}[1]{Sec.~(\ref{#1})}

\def\XXint#1#2#3{{\setbox0=\hbox{$#1{#2#3}{\int}$}
     \vcenter{\hbox{$#2#3$}}\kern-.5\wd0}}

    \newcommand{\beq}{\begin{equation}}
    \newcommand{\eeq}{\end{equation}}
    \newcommand\beqa{\begin{eqnarray}}
    \newcommand\eeqa{\end{eqnarray}}

\title{Resummation of scalar correlator in higher spin black hole background}

\author[a]{Matteo Beccaria } 
\author[a]{, Guido Macorini} 

\affiliation[a]{Dipartimento di Matematica e Fisica Ennio De Giorgi,\\
Universit\`a del Salento \& INFN, Via Arnesano, 73100 Lecce, 
Italy} 
                     
\emailAdd{matteo.beccaria@le.infn.it}
\emailAdd{guido.macorini@le.infn.it}

\abstract{
We consider the proposal that predicts holographic duality between certain 2D minimal models at large central charge
and  Vasiliev 3D higher spin gravity with a single complex field. We compute the scalar correlator in the background of 
a higher spin black hole at order $\mc O(\alpha^{5})$ in the chemical potential $\alpha$ associated with the spin-3 charge.
The calculation is performed at generic values of the symmetry algebra $\hs$ parameter $\lambda$ and for the scalar in
three different representations.  We then study the perturbative data in the large $\lambda$ limit and discover remarkable
regularities. This leads to formulate a closed formula for the resummation of the leading and subleading terms 
that scale like 
$\mc O(\alpha^{n}\lambda^{2n})$ and $\mc O(\alpha^{n}\,\lambda^{2n-1})$ respectively.
}

\allowdisplaybreaks

\begin{document} \maketitle

\bigskip

\section{Introduction}

In the striking paper \cite{Gaberdiel:2010pz}, Vasiliev higher spin theory 
on $AdS_{3}$ coupled to a complex scalar \cite{Vasiliev:1995dn,Vasiliev:1996hn}  has been shown to be 
holographically dual to the conformal 2D  coset conformal minimal models $\mc W_{N}$  at large $N$.
The precise limit that is considered has a large central charge and is parametrised by a single constant $\lambda$
whose gravity meaning is that of labelling a family of $AdS_{3}$ vacua.
The symmetry content of the two sides of the correspondence matches. In particular, the asymptotic symmetry 
algebra of the higher spin gravity theory is $\mc W_{\infty}[\lambda]$ 
\cite{Gaberdiel:2012ku,Henneaux:2010xg,Campoleoni:2010zq,Gaberdiel:2011wb} which is also the classical symmetry 
of the dual CFT \cite{Gaberdiel:2010pz,Gaberdiel:2012ku}.

\medskip
The bulk theory admits higher spin generalisations of the BTZ black hole \cite{Banados:1992wn,Banados:1992gq}
as reviewed in \cite{Ammon:2012wc}. The thermodynamical properties of these very interesting objects are in agreement with 
the proposed duality \cite{Kraus:2011ds,Gaberdiel:2012yb}. However, these tests do not probe the dynamics of the higher spin gravity complex scalar that plays no role in matching the black-hole entropy. Indeed, black holes in $AdS_{3}$ are universal and 
explore the CFT thermodynamics at high temperature, which in two dimensions is only determined by the chiral algebra
and is insensitive to the microscopic features of the conformal theory.

\medskip
The quantum fluctuations of the complex scalar have been studied in \cite{Gutperle:2011kf} by computing 
the bulk-to-boundary scalar correlator in the background of the 3D higher spin black
hole analysed in \cite{Kraus:2011ds}, at least  to first order in the spin-3 charge  chemical potential $\alpha$ and for a special 
value of the $\lambda$ parameter. Later, the authors of \cite{Gaberdiel:2013jca}, extended the calculation to 
$\mc O(\alpha^{2})$ and, remarkably, demonstrated explicitly the agreement with a CFT calculation.

\medskip
In this paper, we considered again the  correlator for the scalar field transforming in three different representations of the higher spin algebra. To be precise, this means that one assigns a higher representation $\Lambda_{+}$ of the symmetry algebra to the scalar dual CFT primary. In the bulk, the Vasiliev master field $C$ obeys 
the equation $dC+A\,C-C\,\overline A=0$ where the higher spin gauge connections $A$, $\overline A$
are in the representation $\Lambda_{+}$. This implies
that, in the bulk, the master field $C$ actually corresponds to an infinite tower of fields with different masses. The details of the spectrum can be worked out by decomposing $\Lambda_{+}$ into irreducible representations of $\mk{sl}(2)$
as discussed in \cite{Hijano:2013fja}. In each case, we computed the propagator at order $\mc O(\alpha^{5})$, by applying  the 
powerful methods introduced in \cite{Kraus:2012uf,Gaberdiel:2013jca}. The reason for such a brute-force calculation
is the idea that some regularity could be observable once we have enough perturbative data to inspect. In particular, 
we looked for special features of the large $\lambda$ regime in the spirit of  \cite{Beccaria:2013dua}.
The idea is that, beyond the large $N$ limit that defines the correspondence, there is another parameter that it is meaningful to take large. Indeed, for $\lambda=-\mc N$, the higher spin calculation truncates to that of a $\mk{sl}(\mc N)\oplus \mk{sl}(\mc N)$ Chern-Simons 3D gravity theory. Taking $\mc N$ to be large could lead to some simplifications in the 
structure of the perturbative corrections as it happens in $SU(\mc N)$ gauge theories. 
This kind of reasoning 
proved to be successful in the study of the black hole partition function  \cite{Beccaria:2013dua}.

\medskip
Although optimistic, this simple attitude turns out to be correct, even beyond the leading order. For the three considered representations of the scalar, we show that all corrections of the order $\mc O(\alpha^{n}\lambda^{2n})$ and $\mc O(\alpha^{n}\lambda^{2n-1})$
can be resummed in closed form. Also, the extension of the results to a generic $K$-antisymmetric representation 
$(\Box^{\otimes K})_{A}$ appears to be straightforward, once one recognises that the representation dependent feature of the resummation formula is encoded in the spin-3 zero mode eigenvalue of the highest 
weight of $(\Box^{\otimes K})_{A}$.

\medskip
The plan of the paper is the following. In \sref{sec:hsbh}, we review the formulation of higher spin black holes
in $\hs\oplus\hs$ Chern-Simons 3D gravity. In \sref{sec:correlator}, we summarise the methods that allow to evaluate 
the scalar correlator. In \sref{sec:structure}, we present the general structure of the correlator in perturbation theory 
and in \sref{sec:calculation} we give some details of the actual $\mc O(\alpha^{5})$ calculation. The results are
non trivially checked in \sref{sec:checks}. Finally, \sref{sec:resum} is devoted to our proposed resummation formula.

\section{Higher spin black hole in $\hs\oplus\hs$ Chern-Simons gravity}
\label{sec:hsbh}

The gauge fields of $\hs\oplus\hs$ Chern-Simons gravity in $AdS_{3}$ are $A$ and $\overline A$, and 
obey the equation of motion
\be
dA+A\wedge \star A = 0,
\ee
as well as  its conjugate. The generators of $\hs$ are usually denoted $V^{s}_{m}$ with $|m|<s$. Space-time is described by a 
radial coordinate $\rho$ and Euclidean torus coordinates $(z, \overline z)$ with periodic identification 
\be
(z, \overline z)\sim(z+2\,\pi, \overline z+2\,\pi)\sim (z+2\,\pi\,\tau, \overline z+2\,\pi\,\overline \tau).
\ee
As shown 
in~\cite{Coussaert:1995zp}, it is possible to choose a gauge where the black hole solution can be written in the form 
\be
\label{eq:gauge-choice}
A(\rho, z, \overline z) = b^{-1}\,a\,b+b^{-1}\,db, \quad
\overline A(\rho, z, \overline z) = b\,\overline a\,b^{-1}+b\,db^{-1},
\ee
where $b=e^{\rho\,V^{2}_{0}}$ and $a, \overline a$ are constant connections without radial components $a_{\rho}=\overline a_{\rho}=0$. The black hole solution with higher spin charges found in \cite{Kraus:2011ds} has the following
explicit expression of the holomorphic connection
\ba
a_{z} &=& V^{2}_{1}-\frac{2\,\pi\,\mc L}{k}\,V^{2}_{-1}-\frac{\pi\,\mc W}{2\,k}\,V^{3}_{-2}
+\sum_{n=4}^{\infty}\mc J_{n}\,V^{n}_{-n+1}, \\
a_{\overline z} &=& -\frac{\alpha}{\overline \tau}\,a_{z}\star a_{z}-\mbox{trace}.
\ea
Here, $\star$ is the {\em lone star product} \cite{Pope:1989sr}, $(\mc L, \mc W)$ are the stress tensor and spin-3 charges, while $\mc J_{n}$ are higher spin charges. Finally, $k$ is the Chern-Simons level. The explicit expansion of the charges
in powers of the chemical potential $\alpha$ associated with the spin-3 charge have been derived in \cite{Kraus:2011ds,Beccaria:2013dua}. At order $\mc O(\alpha^{5})$, we shall need the following expressions,
\ba
\mc L &=& -\frac{k}{8 \left(\pi  \tau ^2\right)}+\frac{\alpha ^2 k \left(\lambda ^2-4\right)}{24 \pi  \tau
   ^6}-\frac{\alpha ^4 \left(k \left(\lambda ^2-7\right) \left(\lambda ^2-4\right)\right)}{24
   \left(\pi  \tau ^{10}\right)}+\mc O\left(\alpha ^6\right), \\
\mc W &=& -\frac{\alpha  k}{3 \left(\pi  \tau ^5\right)}+\frac{10 \alpha ^3 k
   \left(\lambda ^2-7\right)}{27 \pi  \tau ^9}-\frac{\alpha ^5 \left(k
   \left(5 \lambda ^4-85 \lambda ^2+377\right)\right)}{9 \left(\pi  \tau
   ^{13}\right)}+\mc O\left(\alpha ^7\right), \\
\mc J_{4} &=& \frac{7 \alpha ^2}{36 \tau ^8}-\frac{7 \alpha ^4 \left(2 \lambda ^2-21\right)}{36 \tau
   ^{12}}+O\left(\alpha ^6\right), \\
\mc J_{5} &=& \frac{5 \alpha ^3}{18 \tau ^{11}}-\frac{\alpha ^5 \left(44 \lambda
   ^2-635\right)}{54 \tau ^{15}}+\mc O\left(\alpha ^7\right), \\
\mc J_{6} &=& \frac{143 \alpha ^4}{324 \tau ^{14}}+\mc O\left(\alpha ^6\right), \\
\mc J_{7} &=& \frac{182 \alpha ^5}{243 \tau ^{17}}+\mc O\left(\alpha ^7\right).
\ea
From the explicit form of the charges, one can derive the thermal partition function and the black hole entropy 
\cite{deBoer:2013gz,Kraus:2013esi}.

\section{Scalar correlator in higher spin gravity}
\label{sec:correlator}

Let us briefly summarise how to compute the scalar bulk-boundary propagator in 3D higher spin gravity, see
for instance \cite{Kraus:2012uf,Vasiliev:1992gr,Chang:2011mz,Ammon:2011ua,Hijano:2013fja}.
 Our presentation closely follows \cite{Gaberdiel:2013jca} whose notation we adopt.
Assuming the gauge choice (\ref{eq:gauge-choice}), we define
\ba
\Lambda_{0}&=& a_{\mu}\,x^{\mu}, \qquad \Lambda_{\rho} = b^{-1}\star \Lambda_{0}\star b, \\
\overline \Lambda_{0} &=& \overline a_{\mu}\,x^{\mu}, \qquad \overline \Lambda_{\rho} = b\star \overline \Lambda_{0}\star b^{-1}.
\ea
Then, for a bulk scalar with mass $m^{2}=\Delta\,(\Delta-2)$ transforming in a representation of $\hs$, the bulk-boundary
propagator reads
\be
\Phi(z, \overline z, \rho; 0) = e^{\Delta\,\rho}\,\mbox{Tr}\bigg[
e^{-\Lambda_{\rho}}\star c \star e^{\overline \Lambda_{\rho}}\bigg],
\ee
where $c$ is a highest weight of $\hs$, {\em i.e.} an eigenstate of $V^{2}_{0}$ under star product. 
The boundary two-point correlator between two dual fields
at positive infinity can be extracted by taking the $\rho\to+\infty$ limit giving from AdS/CFT duality
\be
\Phi(z, \overline z, \rho; 0) \sim e^{-\Delta\,\rho}\,\langle \overline\varphi(z, \overline z)\,\varphi(0,0)\rangle.
\ee
In the following, we shall consider the defining representation of $\hs$ and its antisymmetric powers.
Then, $c$ is the projector onto the highest weight state $c = |\mbox{hw}\rangle\langle\mbox{hw}|$ that can be built
explicitly from the infinite-dimensional matrix realisation of $\hs$ (see Appendix \ref{app:hs}). Thus,
\be
\Phi(z, \overline z, \rho; 0) = e^{\Delta\,\rho}\,\langle\mbox{hw}| e^{\overline \Lambda_{\rho}}\,e^{-\Lambda_{\rho}}|\mbox{hw}\rangle,
\ee
and in the large $\rho$ limit we find
\be
\langle \overline\varphi(z, \overline z)\,\varphi(0,0)\rangle = \langle -\mbox{hw}|e^{-\Lambda_{0}}|\mbox{hw}\rangle\,\,\langle \mbox{hw}|e^{\overline\Lambda_{0}}|-\mbox{hw}\rangle.
\ee
Due to the above factorisation, we shall consider the purely left-moving part without loosing any information.
In particular, this means that $a_{\overline z}$ will play no role in the following.

\section{Structure of the scalar correlator in the black hole background}
\label{sec:structure}

For a scalar in a generic representation of $\hs$, we have \cite{Kraus:2012uf,Gaberdiel:2013jca}~\footnote{
The sum over images required to impose periodicity $(z,\overline z)\to (z+2\pi,\overline z+2\pi)$ is left implicit.
}
\ba
\langle \overline\varphi(z, \overline z)\,\varphi(0,0)\rangle &=& \left(4\,\tau\,\overline\tau\,\sin\frac{z}{2\,\tau}\,
\sin\frac{\overline z}{2\,\overline \tau}\right)^{-\Delta}\,R(z, \overline z), \\
R(z, \overline z) &=&
1+\sum_{n=1}^{\infty}\frac{\alpha^{n}}{\tau^{2n}}\,R^{(n)}(z, \overline z).
\ea
The corrections $R^{(1)}$ and $R^{(2)}$ have been computed in \cite{Gaberdiel:2013jca} for the fundamental 
and $\sm\yng(1,1)$ representations. The calculation is done in the bulk and it is matched on the CFT side 
where the relevant dual quantity is the torus two-point function of a scalar primary 
in the presence of a deformation of the conformal theory by a holomorphic spin-3 operator. 

Here, we perform an order $\mc O(\alpha^{5})$ calculation including also the 
$\sm\yng(1,1,1)$ representation. In all cases, the general structure of $R^{(n)}$ turns out to be the following:
\be
\label{eq:Ansatz}
R^{(n)}(z, \overline z) = \frac{1}{\sin^{2n}\left(\frac{\mc Z}{2}\right)}\sum_{m=0}^{n} (\mc Z-\overline {\mc Z})^{m}\sum_{k=0}^{n} p^{(n)}_{m,k}(\lambda) \left\{\begin{array}{ll}
\sin(k\,\mc Z), & n+m\ \  \mbox{odd} \\
\cos(k\,\mc Z), & n+m\ \  \mbox{even}
\end{array}\right. ,
\ee
where $p^{(n)}_{m,k}(\lambda)$ are degree $2n$ polynomials and the variables $\mc Z$, $\overline {\mc Z}$ are 
\be
\mc Z = \frac{z}{\tau}, \qquad \overline{\mc Z} = \frac{\overline z}{\overline \tau}.
\ee

\section{Calculation of  $\mc O(\alpha^{5})$ correlators}
\label{sec:calculation}

The extension of the results obtained  in \cite{Gaberdiel:2013jca} is fully straightforward. Here, we just point out a few details that can be useful in order to increase the efficiency of the calculation. First, one assumes that the functions 
$p^{(n)}_{m,k}(\lambda)$ are polynomials. Then, the corrections $R^{(n)}$ are evaluated at $\lambda=-\mc N$ for 
$\mc N = 3, 4, \dots$ up to a point where $p^{(n)}_{m,k}(\lambda)$ can be consistently fixed. We always pushed the calculation some order further in order to confirm that $p^{(n)}_{m,k}(\lambda)$ are actually polynomials.

At $\lambda=-\mc N$, we use the finite-dimensional matrix representation of the generators $V^{s}_{m}$. 
The matrix element $\langle -\mbox{hw}|e^{-\Lambda_{0}}|\mbox{hw} \rangle$ is evaluated for the scalar transforming
in the $K-$antisymmetric representation $(\Box^{\otimes K})_{A}$ using \be
|\mbox{hw}_{K}\rangle = \frac{1}{\sqrt{K!}}|1\rangle\otimes |2\rangle\otimes\cdots\otimes |K\rangle+\mbox{signed permutations},
\ee
and computing
\be
\langle -\mbox{hw}_{K}|e^{-\Lambda_{0}}|\mbox{hw}_{K}\rangle = 
\sum  \langle \mc N|e^{-\Lambda_{0}}|1 \rangle\, \langle \mc N-1|e^{-\Lambda_{0}}|2 \rangle\cdots
 \langle \mc N-k+1|e^{-\Lambda_{0}}|k \rangle,
\ee
where the sum is over the signed permutations of  the labels $\{1, \dots, k\}$ in the kets.

The most time consuming part of the calculation is the construction of the exponential $e^{-\Lambda_{0}}$.
We expand $-\Lambda_{0} = \sum_{n=0}^{\infty}\alpha^{n}\,X_{n}$ and write the expansion of the exponential 
as 
\be
E(t) = e^{t\,\sum_{n=0}^{\infty}\alpha^{n}\,X_{n}} = \sum_{n=0}^{\infty}\alpha^{n}\,E_{n}(t).
\ee
From 
\be
E'(t) = -\Lambda_{0}\,E(t), 
\ee
we obtain 
\ba
E_{0}(t) &=& e^{t\,X_{0}},\\
E_{n}(t) &=& \int_{0}^{t} E_{0}(t-s)\,\sum_{m=1}^{n-1} X_{m}\,E_{n-m}(s)\,ds, \qquad n\ge 1 .
\ea
Finally, the desired exponential is recovered by setting $t=1$. The building block $E_{0}(t)$ is efficiently computed
by noting that $X_{0}$ is diagonalised exploiting the relation~\footnote{
Notice that (\ref{eq:diag}) is consistent with the fact that the BTZ black hole holonomy is in the center of  the group
$HS[\lambda]$ which is the formal exponential of $\hs$ \cite{Kraus:2012uf}. Indeed, this condition requires 
$\exp(2\,\pi\,\tau\,\frac{i}{\tau}V^{2}_{0}) = \exp(2\,\pi\,i\,V^{2}_{0})=\mathbb I$ which is true since $V^{2}_{0}$ is diagonal with integer non zero elements.
}
\be
\label{eq:diag}
e^{-\frac{i}{4\tau}\,V^{2}_{-1}}\,e^{2\,i\,\tau\,V^{2}_{1}}\,\left(V^{2}_{1}+\frac{1}{4\tau^{2}}\,V^{2}_{-1}\right)
e^{-2\,i\,\tau\,V^{2}_{1}}\,e^{\frac{i}{4\tau}\,V^{2}_{-1}} = \frac{i}{\tau}\,V^{2}_{0},
\ee
This change of basis can be done at the beginning and undone before taking matrix elements.

We have computed the 
polynomials $p^{(n)}_{m,k}(\lambda)$ for $n\le 5$ for the the three representations $\sm\yng(1), \sm\yng(1,1), \sm\yng(1,1,1)$. The results for $n\le 4$ are listed in Appendix \ref{app:poly}~\footnote{The explicit polynomials for $n=5$ are available upon request to the authors.}.

\section{Consistency checks}
\label{sec:checks}

\subsection{Representation constraints}

Let us denote by $\wt R$ the ratio $R$ with the replacement $\alpha\to -\alpha$. For simplicity, we can set 
$\overline {\mc Z}=0$, since that variable is always paired with $\mc Z$ according to (\ref{eq:Ansatz}).
We have 
\ba
\mc N=-1\ :  & \qquad R_{\Box}=1, \\
\mc N=-2\ :  & \qquad R_{\Box}=R_{\sm\yng(1,1)}=1, \\
\mc N=-3\ :  & \qquad R_{\sm\yng(1,1,1)}=1, \qquad R_{\Box}=\wt R_{\sm\yng(1,1)}, \\
\mc N=-4\ :  & \qquad R_{\Box}=\wt R_{\sm\yng(1,1,1)}, \\
\mc N=-5\ :  & \qquad R_{\sm\yng(1,1)}=\wt R_{\sm\yng(1,1,1)}.
\ea
These properties are simply understood in terms of equivalences of representations of $\mk{sl}(\mc N)$
 at the above special values of $\mc N$.

\subsection{Zero temperature limit}

An interesting regime is the zero-temperature limit where $\mu=\alpha/\overline \tau$ is fixed and $\tau, \overline \tau\to \infty$. This is the chiral deformation background discussed in \cite{Kraus:2012uf,Ammon:2011ua}. The constant flat connection 
reads in this limit
\be
a = V^{2}_{1}\,dz-\mu\,V^{3}_{2}\,d\overline z, \qquad
\overline a = V^{2}_{-1}\,d\overline z.
\ee
In \cite{Gaberdiel:2013jca}, it has been proved that 
\be
\mathop{\lim_{\tau, \overline \tau\to \infty}}_{\mbox{\tiny fixed}\ \mu} R_{\Box} = \sum_{n=0}^{\infty}
\left(\frac{\mu\,\overline z}{z^{2}}\right)^{n}\,\frac{\Gamma(2n+1+\lambda)}{n!\,\Gamma(1+\lambda)}.
\ee
In Appendix \ref{app:zerotemp}, we prove with a similar calculation that 
\be
\label{eq:zt}
\mathop{\lim_{\tau, \overline \tau\to \infty}}_{\mbox{\tiny fixed}\ \mu} R_{\sm\yng(1,1)} = \sum_{n=0}^{\infty}
\left(\frac{\mu\,\overline z}{z^{2}}\right)^{n}\,
\sum_{m=0}^{n} \frac{(1+\lambda)_{2m}^{+}\,(1+\lambda)_{2\,(n-m)}^{+}}{m!\,(n-m)!}\,\frac{\lambda+n+1+2\,(n-2m)^{2}}{\lambda+1},
\ee
where $(a)_{n}^{+}=a(a+1)\cdots(a+n-1)$ is the ascending Pochammer symbol. The explicit first terms are
\ba
\lefteqn{\mathop{\lim_{\tau, \overline \tau\to \infty}}_{\mbox{\tiny fixed}\ \mu} R_{\sm\yng(1,1)}=} && 
\ \ \ \qquad \qquad 1+\frac{2 (\lambda +2) (\lambda +4) \mu  {\overline z}}{z^2}+\frac{2 (\lambda +2) (\lambda +3)
   \left(\lambda ^2+9 \lambda +23\right) \mu ^2 {\overline z}^2}{z^4}\nonumber \\
   &&+\frac{4 (\lambda +2)
   (\lambda +3) (\lambda +4) (\lambda +6) \left(\lambda ^2+9 \lambda +29\right) \mu ^3
   {\overline z}^3}{3 z^6} \\
   && +\frac{2 (\lambda +2) (\lambda +3) (\lambda +4) (\lambda +5)
   \left(\lambda ^4+22 \lambda ^3+197 \lambda ^2+872 \lambda +1641\right) \mu ^4
   {\overline z}^4}{3 z^8}+\cdots .
   \ea
We have checked that our explicit expression for $R_{\sm\yng(1,1)}$ obeys this limit at order $\mc O(\mu^{5})$.

\section{Leading and subleading order resummation at large $\lambda$}
\label{sec:resum}

An interesting limit of the scalar correlator is the one where $\lambda\to\infty$. This regime has been recently explored
in \cite{Beccaria:2013dua} in the case of the higher spin black hole partition function. One finds that, with a suitable normalisation of $\hs$ generators, the large $\lambda$ limit is smooth and non trivial. Indeed, all higher spin charges are turned on and contribute. Here, we shall also find that the correlators have a sensible large $\lambda$ limit. 
The truncation at $\lambda=-\mc N$ suggest a possible treatment of this regime. The considered observables 
have a polynomial dependence on $\lambda$, at least at each order of perturbation theory in $\alpha$. 
This means that the large $\lambda$ limit in $\hs$ is the same as the large $\mc N$ limit in the 
$\mk{sl}(\mc N)\oplus \mk{sl}(\mc N)$ Chern-Simons gravity. One expects that considering $\mk{sl}(\infty)$ could lead
to some special properties as in the case of standard $1/\mc N$ expansion of $SU(\mc N)$ invariant gauge theories. 

In the present case of the scalar correlator, we have investigated the large $\lambda$ limit and found a remarkable regularities 
of the leading and subleading terms order by order in the expansion in powers of the 
spin-3 charge chemical potential. Indeed, from the explicit expressions of the ratio $R_{r}$, we discover that for the three
considered representations $r=\sm\yng(1), \sm\yng(1,1), \sm \yng(1,1,1)$ it is possible to resum 
all the contributions at leading order $\mc O(\alpha^{n}\,\lambda^{2n})$,
 as well as the next-to-leading terms $\mc O(\alpha^{n}\,\lambda^{2n-1})$. To present the resummation, we introduce
 the following coefficient associated with the $K$-antisymmetric representation 
 \be
 c_{r=(\Box^{\otimes K})_{A}} =  \frac{K}{6}(\lambda+K)(\lambda+2K),
 \ee
 and the two functions
\ba
f_{\rm LO}(\mc Z, \overline{\mc Z}) &=& \frac{3\,\sin\mc Z-(2+\cos\mc Z)\,(\mc Z-\overline{\mc Z})}{\sin^{2}\frac{\mc Z}{2}}, \\
f_{\rm NLO}(\mc Z, \overline{\mc Z}) &=& 6-4\,\cos\mc Z-2\cos(2\mc Z)-(10\,\sin\mc Z+\sin(2\mc Z))\,(\mc Z-\overline{\mc Z})\nonumber \\
&&+3\,(1+\cos\mc Z)\,(\mc Z-\overline{\mc Z})^{2}.
\ea
Then, we have the rather simple formula
\be
\label{eq:resummation}
R_{r} \stackrel{\lambda\to\infty}{=} \exp\left(
c_{r}\,f_{\rm LO}\,\frac{\alpha}{2\,\tau^{2}}\right)\,\left(1+c_{r}\,\frac{f_{\rm NLO}}{\sin^{4}\frac{\mc Z}{2}}
\,\frac{\lambda\,\alpha^{2}}{8\,\tau^{4}}
\right)+\mc O(\alpha^{n}\,\lambda^{2n-2}).
\ee
Resummation works as follows. The argument of the exponential captures all terms $\sim \alpha^{n}\lambda^{2n}$ and 
introduces also some contributions of order $\alpha^{n}\lambda^{2n-1}$. The  
correction in parenthesis completes the resummation of all contributions of order $\alpha^{n}\lambda^{2n-1}$.
From the CFT calculation of the $\mc O(\alpha)$ contribution to $R$, we identify the meaning of $c_{r}$. It is 
the eigenvalue of the  zero mode
of the $\mc W_{\infty}$ spin-3 generator on the dual primary. In terms of $\hs$, this is the eigenvalue of 
 $V^{3}_{0}$ on the highest weight. A straightforward calculation gives indeed 
  \be
 V^{3}_{0}\,|\mbox{hw}_{K}\rangle = \frac{K}{6}(\lambda+K)(\lambda+2K)\,|\mbox{hw}_{K}\rangle.
 \ee
 The resummation formula (\ref{eq:resummation}) is our conjecture and we have tested it up to the order $\mc O(\alpha^{5})$ included.

\section{Conclusions}

In this paper, we have considered the duality between Vasiliev 3D higher spin gravity and $\mc W_{N}$ minimal models.
We have computed the correlator of the bulk complex scalar in the background of a higher spin black hole and we have determined the exact  dependence on the symmetry parameter $\lambda$ at order $\mc O(\alpha^{5})$ where $\alpha$ is the chemical potential associated with the spin-3 charge of the black hole.
We have analysed the perturbative data in the large $\lambda$ limit where we had some expectations that some simplification could occur. We found special regularities and proposed a resummation of all the leading and subleading contributions that take the form 
$\mc O(\alpha^{n}\cdot\{\lambda^{2n}, \lambda^{2n-1}\})$. 

We believe that such non-trivial property of the perturbative expansion could be explained,  and possibly extended, in terms
of a systematic $1/\mc N$ expansion of the $\mk{sl}(\mc N)\oplus \mk{sl}(\mc N)$ Chern-Simons gravity theory.
Alternatively, a suitable contraction of the symmetry algebra $\hs$ could be devised in order to capture the corrections considered in this paper.

%

\appendix

\section{The infinite dimensional algebra $\hs$}
\label{app:hs}

The $\hs$ algebra is generated by  $V^{s}_{m}$ with 
$s\ge 2$ and $|m|<s$. The lone-star operation \cite{Pope:1989sr} is the associative product defined by 
\be
V^{s}_{m}\star V^{t}_{n} = \frac{1}{2}\sum_{u=1}^{s+t-1} g^{st}_{u}(m,n,\lambda)\,V^{s+t-u}_{m+n},
\ee
with 
\ba
g^{st}_{u}(m,n,\lambda) &=& \frac{(1/4)^{u-2}}{2(u-1)!}{}_{4}F_{3}\left(
\left.
\begin{array}{cc}
\frac{1}{2}+\lambda, \ \frac{1}{2}-\lambda, \ \frac{2-u}{2}, \ \frac{1-u}{2} \\
\frac{3}{2}-s, \ \frac{3}{2}-t, \ \frac{1}{2}+s+t-u
\end{array}\right | 1
\right)\times  \\
&& \sum_{k=0}^{u-1}(-1)^{k}\binom{u-1}{k}(s-1+m)_{u-1-k}^{-}\,(s-1-m)_{k}^{-}(t-1+n)_{k}^{-}(t-1-n)_{u-1-k}^{-},
\nonumber
\ea
where $(a)_{n}^{-} = a(a-1)\cdots(a-n+1)$ is the descending Pochammer symbol. An infinite-dimensional matrix representation of $\hs$ is obtained by starting with 
the following infinite-dimensional matrix representation of $\mk{sl}(2)$ (we list the non-zero elements )
\ba
\left(V^{2}_{0}\right)_{n,n} &=& \frac{-\lambda+1}{2}-n, \\
\left(V^{2}_{1}\right)_{n+1,n} &=& -\sqrt{(-\lambda-n)\,n}, \\
\left(V^{2}_{-1}\right)_{n,n+1} &=& \sqrt{(-\lambda-n)\,n}, 
\ea
and building the other generators according to 
\be
V^{s}_{m}=(-1)^{s-1-m}\,\frac{(s+m-1)!}{(2s-2)!}\,\mbox{Ad}_{V^{2}_{-1}}^{s-m-1}\left(V^{2}_{1}\right)^{s-1}.
\ee
A very important property of this representation is that setting $\lambda=-\mc N$, a negative integer, and 
restricting $V^{s}_{m}$ to the set with $s\le \mc N$, we obtain a $\mc N\times \mc N$ dimensional representation
of $\mk{sl}(\mc N)$ by restricting the infinite dimensional matrices to their first $\mc N$ rows and columns.
Also, the lone-star product becomes simply matrix multiplication.

\section{Non-zero polynomials $p^{(n)}_{m,k}(\lambda)$ for $n\le 4$}
\label{app:poly}

In this section we report the explicit expressions for the polynomials appearing in (\ref{eq:Ansatz}).

\subsection{Fundamental representation}

Let us define 
\begin{equation}
\mathit{f}=(\lambda +1) (\lambda +2).
\end{equation}
We have:
{\footnotesize

\begin{eqnarray}
&p^{(1)}_{0,1}(\lambda)=& \frac{\mathit{f}}{4} \nonumber\\
&p^{(1)}_{1,0}(\lambda)=& -\frac{\mathit{f}}{6}\nonumber \\
&p^{(1)}_{1,1}(\lambda)=& -\frac{\mathit{f}}{12} 
\end{eqnarray}

\begin{eqnarray}
&p^{(2)}_{0,0}(\lambda)=&  \frac{\mathit{f}}{64}  (\lambda +4) (\lambda +7)  \nonumber\\
&p^{(2)}_{0,1}(\lambda)=&  -\frac{\mathit{f}}{12}  (\lambda +4)  \nonumber\\
&p^{(2)}_{0,2}(\lambda)=&  -\frac{\mathit{f}}{192}  (\lambda +4) (3 \lambda +5) \nonumber \\
&p^{(2)}_{1,1}(\lambda)=&  -\frac{\mathit{f}}{24}  (\lambda +4)^2  \nonumber\\
&p^{(2)}_{1,2}(\lambda)=&  -\frac{\mathit{f}}{96} (\lambda +1) (\lambda +4) \nonumber \\
&p^{(2)}_{2,0}(\lambda)=&  \frac{\mathit{f}}{64}  \left(\lambda ^2+7 \lambda +14\right) \nonumber \\
&p^{(2)}_{2,1}(\lambda)=&  \frac{\mathit{f}}{144}  (\lambda +2) (2 \lambda +11)  \nonumber\\
&p^{(2)}_{2,2}(\lambda)=&  \frac{\mathit{f}}{576}  (\lambda +1) (\lambda +2) 
\end{eqnarray}

\begin{eqnarray}
 &p^{(3)}_{0,1}(\lambda)=& \frac{\mathit{f}}{1536} (\lambda +3) (\lambda +4) (\lambda +6) (3 \lambda +35) \nonumber\\
 &p^{(3)}_{0,2}(\lambda)=& -\frac{\mathit{f}}{96}  (\lambda +3) (\lambda +4) (\lambda +6) \nonumber\\
 &p^{(3)}_{0,3}(\lambda)=& -\frac{\mathit{f}}{1536} (\lambda +1) (\lambda +3) (\lambda +4) (\lambda +6) \nonumber\\
 &p^{(3)}_{1,0}(\lambda)=& -\frac{\mathit{f}}{1152} \left(3 \lambda ^4+68 \lambda ^3+539 \lambda ^2+1942 \lambda +2488\right) \nonumber\\
 &p^{(3)}_{1,1}(\lambda)=& -\frac{\mathit{f}}{1536} \left(\lambda ^4+2 \lambda ^3-89 \lambda ^2-842 \lambda -1592\right) \nonumber\\
 &p^{(3)}_{1,2}(\lambda)=& \frac{\mathit{f}}{384}  \left(\lambda ^4+20 \lambda ^3+145 \lambda ^2+418 \lambda +424\right)\nonumber \\
 &p^{(3)}_{1,3}(\lambda)=& \frac{\mathit{f}}{4608} \left(3 \lambda ^4+38 \lambda ^3+149 \lambda ^2+226 \lambda +88\right)\nonumber \\
 &p^{(3)}_{2,1}(\lambda)=& \frac{\mathit{f}}{4608} (\lambda +6) \left(17 \lambda ^3+236 \lambda ^2+1093 \lambda +1594\right) \nonumber\\
 &p^{(3)}_{2,2}(\lambda)=& \frac{\mathit{f}}{1152} (\lambda +2) (\lambda +6) (\lambda +7) (2 \lambda +5)\nonumber \\
 &p^{(3)}_{2,3}(\lambda)=& \frac{\mathit{f}^2}{4608} (\lambda +2) (\lambda +6)\nonumber \\
 &p^{(3)}_{3,0}(\lambda)=& -\frac{\mathit{f}}{10368} \left(11 \lambda ^4+201 \lambda ^3+1412 \lambda ^2+4344 \lambda +4904\right) \nonumber\\
 &p^{(3)}_{3,1}(\lambda)=& -\frac{\mathit{f} }{13824}\left(17 \lambda ^4+318 \lambda ^3+2093 \lambda ^2+5892 \lambda +5972\right) \nonumber\\
 &p^{(3)}_{3,2}(\lambda)=& -\frac{\mathit{f}}{3456} (\lambda +2)^2 (\lambda +4) (\lambda +7) \nonumber\\
 &p^{(3)}_{3,3}(\lambda)=& -\frac{\mathit{f}^3}{41472} 
\end{eqnarray}

\begin{eqnarray}
&p^{(4)}_{0,0}(\lambda)=& \frac{\mathit{f} }{147456}(\lambda +8) \left(9 \lambda ^5+345 \lambda ^4+4925 \lambda ^3+32135 \lambda ^2+100546 \lambda +117400\right) \nonumber\\
&p^{(4)}_{0,1}(\lambda)=& -\frac{\mathit{f}}{4608} (\lambda +8) \left(3 \lambda ^4+82 \lambda ^3+721 \lambda ^2+2762 \lambda +3656\right) \nonumber\\
&p^{(4)}_{0,2}(\lambda)=& -\frac{\mathit{f}}{36864} (\lambda +8) \left(3 \lambda ^5+99 \lambda ^4+1055 \lambda ^3+4925 \lambda ^2+9526 \lambda +5896\right) \nonumber\\
&p^{(4)}_{0,3}(\lambda)=& \frac{\mathit{f}}{4608} (\lambda +8) \left(3 \lambda ^4+50 \lambda ^3+305 \lambda ^2+778 \lambda +712\right) \nonumber\\
&p^{(4)}_{0,4}(\lambda)=& \frac{\mathit{f}}{147456} (\lambda +8) \left(3 \lambda ^5+51 \lambda ^4+319 \lambda ^3+877 \lambda ^2+1046 \lambda +392\right)\nonumber \\
&p^{(4)}_{1,1}(\lambda)=& -\frac{\mathit{f}}{9216} \left(3 \lambda ^6+124 \lambda ^5+2013 \lambda ^4+16904 \lambda ^3+76708 \lambda ^2+177768 \lambda +163952\right) \nonumber\\
&p^{(4)}_{1,2}(\lambda)=& -\frac{\mathit{f}}{18432} \left(\lambda ^6+7 \lambda ^5-287 \lambda ^4-5551 \lambda ^3-37282 \lambda ^2-109368 \lambda -117664\right) \nonumber\\
&p^{(4)}_{1,3}(\lambda)=& \frac{\mathit{f}}{9216} (\lambda +2) \left(\lambda ^5+34 \lambda ^4+443 \lambda ^3+2642 \lambda ^2+7368 \lambda +7656\right) \nonumber\\
&p^{(4)}_{1,4}(\lambda)=& \frac{\mathit{f}}{36864} (\lambda +1) \left(\lambda ^5+22 \lambda ^4+171 \lambda ^3+598 \lambda ^2+872 \lambda +352\right) \nonumber\\
&p^{(4)}_{2,0}(\lambda)=& \frac{\mathit{f}}{221184} \left(51 \lambda ^6+1955 \lambda ^5+30137 \lambda ^4+245921 \lambda ^3+1101812 \lambda ^2+2564444 \lambda +2406080\right) \nonumber\\
&p^{(4)}_{2,1}(\lambda)=& \frac{\mathit{f}}{55296} \left(6 \lambda ^6+169 \lambda ^5+1870 \lambda ^4+7807 \lambda ^3+1672 \lambda ^2-73748 \lambda -137360\right) \nonumber\\
&p^{(4)}_{2,2}(\lambda)=& -\frac{\mathit{f}}{27648} \left(6 \lambda ^6+229 \lambda ^5+3541 \lambda ^4+27937 \lambda ^3+118309 \lambda ^2+253066 \lambda +213808\right) \nonumber\\
&p^{(4)}_{2,3}(\lambda)=& -\frac{\mathit{f}}{55296} (\lambda +2) \left(6 \lambda ^5+173 \lambda ^4+1876 \lambda ^3+9351 \lambda ^2+21338 \lambda +18232\right) \nonumber\\
&p^{(4)}_{2,4}(\lambda)=& -\frac{\mathit{f}^2}{221184} \left(3 \lambda ^4+50 \lambda ^3+245 \lambda ^2+406 \lambda +160\right) \nonumber\\
&p^{(4)}_{3,1}(\lambda)=& -\frac{\mathit{f}}{82944} (\lambda +8) \left(19 \lambda ^5+529 \lambda ^4+5869 \lambda ^3+31759 \lambda ^2+82816 \lambda +82876\right) \nonumber\\
&p^{(4)}_{3,2}(\lambda)=& -\frac{\mathit{f}}{165888} (\lambda +8) \left(25 \lambda ^5+643 \lambda ^4+6199 \lambda ^3+28093 \lambda ^2+60556 \lambda +50140\right)\nonumber \\
&p^{(4)}_{3,3}(\lambda)=& -\frac{\mathit{f}}{27648} (\lambda +2)^2 (\lambda +8) \left(\lambda ^3+15 \lambda ^2+63 \lambda +85\right) \nonumber\\
&p^{(4)}_{3,4}(\lambda)=& -\frac{\mathit{f}^3}{331776} (\lambda +1) (\lambda +8) \nonumber\\
&p^{(4)}_{4,0}(\lambda)=& \frac{\mathit{f}}{3981312} \left(227 \lambda ^6+7659 \lambda ^5+107715 \lambda ^4+795357 \lambda ^3+3235542 \lambda ^2+6848460 \lambda +5873560\right) \nonumber\\
&p^{(4)}_{4,1}(\lambda)=& \frac{\mathit{f}}{497664} \left(38 \lambda ^6+1287 \lambda ^5+17562 \lambda ^4+124461 \lambda ^3+482406 \lambda ^2+968940 \lambda +787624\right) \nonumber\\
&p^{(4)}_{4,2}(\lambda)=& \frac{\mathit{f}}{995328} \left(25 \lambda ^6+765 \lambda ^5+9249 \lambda ^4+56655 \lambda ^3+187086 \lambda ^2+317556 \lambda +217496\right)\nonumber \\
&p^{(4)}_{4,3}(\lambda)=& \frac{\mathit{f}}{497664} (\lambda +2)^3 \left(2 \lambda ^3+33 \lambda ^2+168 \lambda +299\right) \nonumber\\
&p^{(4)}_{4,4}(\lambda)=& \frac{\mathit{f}^4}{3981312} 
\end{eqnarray}
}

\subsection{ 2-Antisymmetric representation}
Let us define
\begin{equation}
\mathit{f}=(\lambda +2) (\lambda +4).
\end{equation}
We have:
{\footnotesize

\begin{eqnarray}
&p^{(1)}_{0,1}(\lambda)=& \frac{\mathit{f}}{2} \nonumber\\
&p^{(1)}_{1,0}(\lambda)=& -\frac{\mathit{f}}{3}\nonumber \\
&p^{(1)}_{1,1}(\lambda)=& -\frac{\mathit{f}}{6} 
\end{eqnarray}

\begin{eqnarray}
 &p^{(2)}_{0,0}(\lambda)=&\frac{1}{16} (\lambda +2) \left(\lambda ^3+14 \lambda ^2+74 \lambda +121\right) \nonumber\\
 &p^{(2)}_{0,1}(\lambda)=&-\frac{1}{6} (\lambda +2) \left(\lambda ^2+12 \lambda +26\right) \nonumber\\
 &p^{(2)}_{0,2}(\lambda)=&-\frac{1}{48} (\lambda +2) (\lambda +5) \left(3 \lambda ^2+19 \lambda +31\right) \nonumber\\
 &p^{(2)}_{1,1}(\lambda)=&-\frac{1}{12} (\lambda +2) \left(2 \lambda ^3+25 \lambda ^2+112 \lambda +164\right)\nonumber \\
 &p^{(2)}_{1,2}(\lambda)=&-\frac{1}{24} (\lambda +2) \left(\lambda ^3+11 \lambda ^2+38 \lambda +43\right) \nonumber\\
 &p^{(2)}_{2,0}(\lambda)=&\frac{1}{16} (\lambda +2) \left(\lambda ^3+12 \lambda ^2+52 \lambda +74\right) \nonumber\\
 &p^{(2)}_{2,1}(\lambda)=&\frac{\mathit{f}}{72}  (\lambda +4) (4 \lambda +17) \nonumber\\
 &p^{(2)}_{2,2}(\lambda)=&\frac{\mathit{f}^2}{144} 
\end{eqnarray}

\begin{eqnarray}
&p^{(3)}_{0,1}(\lambda)=& \frac{\mathit{f}}{192}  (\lambda +3) (\lambda +6) \left(3 \lambda ^2+37 \lambda +167\right) \nonumber\\
&p^{(3)}_{0,2}(\lambda)=& -\frac{\mathit{f}}{24}  (\lambda +3) (\lambda +6) (\lambda +8) \nonumber\\
&p^{(3)}_{0,3}(\lambda)=& -\frac{\mathit{f}}{192}  (\lambda +3) (\lambda +6) \left(\lambda ^2+7 \lambda +13\right)\nonumber \\
&p^{(3)}_{1,0}(\lambda)=& -\frac{\mathit{f}}{144}  \left(3 \lambda ^4+61 \lambda ^3+503 \lambda ^2+1979 \lambda +2794\right)\nonumber \\
&p^{(3)}_{1,1}(\lambda)=& -\frac{\mathit{f} }{192} \left(\lambda ^4+10 \lambda ^3-8 \lambda ^2-457 \lambda -1070\right) \nonumber\\
&p^{(3)}_{1,2}(\lambda)=& \frac{\mathit{f}}{48}  \left(\lambda ^4+19 \lambda ^3+145 \lambda ^2+497 \lambda +622\right) \nonumber\\
&p^{(3)}_{1,3}(\lambda)=& \frac{\mathit{f} }{576} \left(3 \lambda ^4+46 \lambda ^3+248 \lambda ^2+581 \lambda +502\right)\nonumber \\
&p^{(3)}_{2,1}(\lambda)=& \frac{\mathit{f}}{576}  \left(17 \lambda ^4+322 \lambda ^3+2488 \lambda ^2+9026 \lambda +11988\right) \nonumber\\
&p^{(3)}_{2,2}(\lambda)=& \frac{\mathit{f}}{288} (\lambda +5) \left(4 \lambda ^3+51 \lambda ^2+224 \lambda +324\right) \nonumber\\
&p^{(3)}_{2,3}(\lambda)=& \frac{\mathit{f}}{576}  (\lambda +2) \left(\lambda ^3+12 \lambda ^2+44 \lambda +54\right) \nonumber\\
&p^{(3)}_{3,0}(\lambda)=& -\frac{\mathit{f} \left(22 \lambda ^4+399 \lambda ^3+2953 \lambda ^2+10212 \lambda +13072\right)}{2592} \nonumber\\
&p^{(3)}_{3,1}(\lambda)=& -\frac{\mathit{f} \left(17 \lambda ^4+312 \lambda ^3+2216 \lambda ^2+7284 \lambda +8936\right)}{1728}\nonumber \\
&p^{(3)}_{3,2}(\lambda)=& -\frac{1}{864} \mathit{f} (\lambda +4)^2 (\lambda +5) (2 \lambda +7) \nonumber\\
&p^{(3)}_{3,3}(\lambda)=& -\frac{\mathit{f}^3}{5184} 
\end{eqnarray}

\begin{eqnarray}
 &p^{(4)}_{0,0}(\lambda)=&\frac{(\lambda +2) \left(9 \lambda ^7+366 \lambda ^6+6632 \lambda ^5+68870 \lambda ^4+439976 \lambda ^3+1700624 \lambda ^2+3588743 \lambda
   +3127340\right)}{9216} \nonumber\\
 &p^{(4)}_{0,1}(\lambda)=&-\frac{1}{576} (\lambda +2) \left(3 \lambda ^6+110 \lambda ^5+1712 \lambda ^4+14507 \lambda ^3+68774 \lambda ^2+167078 \lambda +159968\right) \nonumber\\
 &p^{(4)}_{0,2}(\lambda)=& -\frac{(\lambda +2) \left(3 \lambda ^7+114 \lambda ^6+1880 \lambda ^5+17234 \lambda ^4+93488 \lambda ^3+296192 \lambda ^2+505949 \lambda +360356\right)}{2304}\nonumber
   \\
 &p^{(4)}_{0,3}(\lambda)=&\frac{1}{576} (\lambda +2) \left(3 \lambda ^6+94 \lambda ^5+1216 \lambda ^4+8299 \lambda ^3+31366 \lambda ^2+62182 \lambda +50464\right) \nonumber\\
 &p^{(4)}_{0,4}(\lambda)=&\frac{(\lambda +2) \left(3 \lambda ^7+90 \lambda ^6+1144 \lambda ^5+8002 \lambda ^4+33304 \lambda ^3+82672 \lambda ^2+113389 \lambda +66148\right)}{9216}\nonumber \\
 &p^{(4)}_{1,1}(\lambda)=&-\frac{(\lambda +2) (\lambda +8) \left(6 \lambda ^6+181 \lambda ^5+2414 \lambda ^4+18004 \lambda ^3+77105 \lambda ^2+174640 \lambda +160172\right)}{1152} \nonumber\\
 &p^{(4)}_{1,2}(\lambda)=&-\frac{(\lambda +2) \left(\lambda ^7+21 \lambda ^6+78 \lambda ^5-1894 \lambda ^4-26049 \lambda ^3-139840 \lambda ^2-349987 \lambda -335548\right)}{1152} \nonumber\\
 &p^{(4)}_{1,3}(\lambda)=&\frac{(\lambda +2) \left(2 \lambda ^7+71 \lambda ^6+1106 \lambda ^5+9644 \lambda ^4+50411 \lambda ^3+156968 \lambda ^2+268548 \lambda +194208\right)}{1152} \nonumber\\
 &p^{(4)}_{1,4}(\lambda)=&\frac{(\lambda +2) \left(\lambda ^7+29 \lambda ^6+350 \lambda ^5+2298 \lambda ^4+8903 \lambda ^3+20448 \lambda ^2+25837 \lambda +13828\right)}{2304} \nonumber\\
 &p^{(4)}_{2,0}(\lambda)=&\frac{(\lambda +2) \left(51 \lambda ^7+1870 \lambda ^6+30420 \lambda ^5+285351 \lambda ^4+1653324 \lambda ^3+5831100 \lambda ^2+11379570 \lambda
   +9326444\right)}{13824} \nonumber\\
 &p^{(4)}_{2,1}(\lambda)=&\frac{(\lambda +2) \left(12 \lambda ^7+379 \lambda ^6+5118 \lambda ^5+36678 \lambda ^4+142599 \lambda ^3+260448 \lambda ^2+86208 \lambda -213256\right)}{6912}\nonumber\\
 &p^{(4)}_{2,2}(\lambda)=&-\frac{(\lambda +2) \left(12 \lambda ^7+439 \lambda ^6+7134 \lambda ^5+66189 \lambda ^4+375123 \lambda ^3+1281228 \lambda ^2+2410755 \lambda
   +1907840\right)}{3456} \nonumber\\
 &p^{(4)}_{2,3}(\lambda)=&-\frac{(\lambda +2) \left(12 \lambda ^7+395 \lambda ^6+5598 \lambda ^5+44166 \lambda ^4+208983 \lambda ^3+592416 \lambda ^2+930144 \lambda +622456\right)}{6912}\nonumber\\
 &p^{(4)}_{2,4}(\lambda)=&-\frac{(\lambda +2)^2 \left(\lambda ^3+14 \lambda ^2+56 \lambda +70\right) \left(3 \lambda ^3+34 \lambda ^2+128 \lambda +169\right)}{13824} \nonumber\\
 &p^{(4)}_{3,1}(\lambda)=&-\frac{\mathit{f} \left(38 \lambda ^6+1194 \lambda ^5+16485 \lambda ^4+127347 \lambda ^3+574173 \lambda ^2+1399308 \lambda +1395856\right)}{10368} \nonumber\\
 &p^{(4)}_{3,2}(\lambda)=&-\frac{\mathit{f} \left(25 \lambda ^6+759 \lambda ^5+9822 \lambda ^4+69519 \lambda ^3+282600 \lambda ^2+618708 \lambda +560402\right)}{10368} \nonumber\\
 &p^{(4)}_{3,3}(\lambda)=&-\frac{\mathit{f} (\lambda +4) \left(2 \lambda ^5+46 \lambda ^4+419 \lambda ^3+1897 \lambda ^2+4295 \lambda +3880\right)}{3456} \nonumber\\
 &p^{(4)}_{3,4}(\lambda)=&-\frac{\mathit{f}^2 (\lambda +2) \left(\lambda ^3+13 \lambda ^2+50 \lambda +65\right)}{20736} \nonumber\\
 &p^{(4)}_{4,0}(\lambda)=&\frac{(\lambda +2) \left(227 \lambda ^7+7802 \lambda ^6+119208 \lambda ^5+1045128 \lambda ^4+5645904 \lambda ^3+18594528 \lambda ^2+34117144 \lambda
   +26560720\right)}{248832} \nonumber\\
 &p^{(4)}_{4,1}(\lambda)=&\frac{(\lambda +2) \left(152 \lambda ^7+5234 \lambda ^6+78852 \lambda ^5+676263 \lambda ^4+3558120 \lambda ^3+11403312 \lambda ^2+20392768 \lambda
   +15518560\right)}{124416} \nonumber\\
 &p^{(4)}_{4,2}(\lambda)=&\frac{(\lambda +2) \left(25 \lambda ^7+820 \lambda ^6+11607 \lambda ^5+92220 \lambda ^4+444648 \lambda ^3+1298208 \lambda ^2+2115848 \lambda
   +1476692\right)}{62208} \nonumber\\
 &p^{(4)}_{4,3}(\lambda)=&\frac{\mathit{f} (\lambda +4)^3 \left(8 \lambda ^3+102 \lambda ^2+420 \lambda +577\right)}{124416} \nonumber\\
 &p^{(4)}_{4,4}(\lambda)=&\frac{\mathit{f}^4}{248832} 
\end{eqnarray}
}

\subsection{3-Antisymmetric representation}
Let us define
\begin{equation}
\mathit{f}=(\lambda +2) (\lambda +4).
\end{equation}
We have:
{\footnotesize

\begin{eqnarray}
 &p^{(1)}_{0,1}(\lambda)=& \frac{3 \mathit{f}}{4} \nonumber\\
 &p^{(1)}_{1,0}(\lambda)=& -\frac{\mathit{f}}{2} \nonumber\\
 &p^{(1)}_{1,1}(\lambda)=& -\frac{\mathit{f}}{4} 
\end{eqnarray}

\begin{eqnarray}
 &p^{(2)}_{0,0}(\lambda)=&\frac{3}{64} (\lambda +3) (\lambda +4) \left(3 \lambda ^2+41 \lambda +178\right) \nonumber\\
 &p^{(2)}_{0,1}(\lambda)=&-\frac{1}{4} (\lambda +3) (\lambda +4) (\lambda +14) \nonumber\\
 &p^{(2)}_{0,2}(\lambda)=&-\frac{1}{64} (\lambda +3) (\lambda +4) (\lambda +5) (9 \lambda +62) \nonumber\\
 &p^{(2)}_{1,1}(\lambda)=&-\frac{1}{8} (\lambda +3) (\lambda +4) \left(3 \lambda ^2+38 \lambda +136\right) \nonumber\\
 &p^{(2)}_{1,2}(\lambda)=&-\frac{1}{32} (\lambda +3) (\lambda +4) \left(3 \lambda ^2+35 \lambda +94\right) \nonumber\\
 &p^{(2)}_{2,0}(\lambda)=&\frac{3}{64} (\lambda +3) \left(3 \lambda ^3+49 \lambda ^2+276 \lambda +508\right) \nonumber\\
 &p^{(2)}_{2,1}(\lambda)=&\frac{\mathit{f}}{16}  (\lambda +6) (2 \lambda +9) \nonumber\\
 &p^{(2)}_{2,2}(\lambda)=&\frac{\mathit{f}^2}{64} 
\end{eqnarray}

\begin{eqnarray}
&p^{(3)}_{0,1}(\lambda)=& \frac{3\mathit{f}}{512}  (\lambda +4) \left(9 \lambda ^3+182 \lambda ^2+1419 \lambda +3822\right) \nonumber\\
&p^{(3)}_{0,2}(\lambda)=& -\frac{3\mathit{f}}{32}  (\lambda +4) \left(\lambda ^2+19 \lambda +74\right) \nonumber\\
&p^{(3)}_{0,3}(\lambda)=& -\frac{\mathit{f}}{512}  (\lambda +4) \left(9 \lambda ^3+150 \lambda ^2+811 \lambda +1454\right) \nonumber\\
&p^{(3)}_{1,0}(\lambda)=& -\frac{\mathit{f}}{128}  \left(9 \lambda ^4+212 \lambda ^3+2009 \lambda ^2+8758 \lambda +14072\right) \nonumber\\
&p^{(3)}_{1,1}(\lambda)=& -\frac{\mathit{f}}{512}  \left(9 \lambda ^4+150 \lambda ^3+583 \lambda ^2-1662 \lambda -8200\right) \nonumber\\
&p^{(3)}_{1,2}(\lambda)=& \frac{\mathit{f}}{128}  \left(9 \lambda ^4+204 \lambda ^3+1825 \lambda ^2+7302 \lambda +10808\right) \nonumber\\
&p^{(3)}_{1,3}(\lambda)=& \frac{\mathit{f}}{512}  \left(9 \lambda ^4+182 \lambda ^3+1319 \lambda ^2+4162 \lambda +4856\right) \nonumber\\
&p^{(3)}_{2,1}(\lambda)=& \frac{\mathit{f}}{512}  \left(51 \lambda ^4+1154 \lambda ^3+10399 \lambda ^2+42712 \lambda +65204\right) \nonumber\\
&p^{(3)}_{2,2}(\lambda)=& \frac{\mathit{f}}{128} (\lambda +5) \left(6 \lambda ^3+101 \lambda ^2+576 \lambda +1084\right) \nonumber\\
&p^{(3)}_{2,3}(\lambda)=& \frac{\mathit{f}}{512} (\lambda +3) \left(3 \lambda ^3+49 \lambda ^2+252 \lambda +428\right) \nonumber\\
&p^{(3)}_{3,0}(\lambda)=& -\frac{\mathit{f}}{384}  \left(11 \lambda ^4+243 \lambda ^3+2124 \lambda ^2+8424 \lambda +12456\right) \nonumber\\
&p^{(3)}_{3,1}(\lambda)=& -\frac{\mathit{f}}{512}  \left(17 \lambda ^4+378 \lambda ^3+3221 \lambda ^2+12396 \lambda +17876\right) \nonumber\\
&p^{(3)}_{3,2}(\lambda)=& -\frac{\mathit{f}}{128}  (\lambda +4) (\lambda +5) (\lambda +6)^2 \nonumber\\
&p^{(3)}_{3,3}(\lambda)=& -\frac{\mathit{f}^3}{1536} 
\end{eqnarray}

\begin{eqnarray}
&p^{(4)}_{0,0}(\lambda)=& \frac{(\lambda +3) \left(243 \lambda ^7+11043 \lambda ^6+224709 \lambda ^5+2623025 \lambda ^4+18820392 \lambda ^3+82210172 \lambda ^2+199702896 \lambda
   +205364480\right)}{49152} \nonumber\\
&p^{(4)}_{0,1}(\lambda)=& -\frac{(\lambda +3) \left(27 \lambda ^6+1200 \lambda ^5+22213 \lambda ^4+221052 \lambda ^3+1232956 \lambda ^2+3587088 \lambda +4202752\right)}{1536} \nonumber\\
&p^{(4)}_{0,2}(\lambda)=& -\frac{(\lambda +3) \left(81 \lambda ^7+3537 \lambda ^6+68055 \lambda ^5+736907 \lambda ^4+4786296 \lambda ^3+18477716 \lambda ^2+39096528 \lambda
   +34934528\right)}{12288} \nonumber\\
&p^{(4)}_{0,3}(\lambda)=& \frac{(\lambda +3) \left(27 \lambda ^6+1104 \lambda ^5+18149 \lambda ^4+155004 \lambda ^3+729404 \lambda ^2+1799952 \lambda +1822976\right)}{1536} \nonumber\\
&p^{(4)}_{0,4}(\lambda)=& \frac{(\lambda +3) \left(81 \lambda ^7+3105 \lambda ^6+50583 \lambda ^5+454651 \lambda ^4+2438328 \lambda ^3+7814356 \lambda ^2+13871568 \lambda
   +10526464\right)}{49152} \nonumber\\
&p^{(4)}_{1,1}(\lambda)=& -\frac{(\lambda +3) \left(81 \lambda ^7+3546 \lambda ^6+68847 \lambda ^5+763034 \lambda ^4+5173692 \lambda ^3+21283280 \lambda ^2+48737472 \lambda
   +47501024\right)}{3072} \nonumber\\
&p^{(4)}_{1,2}(\lambda)=& -\frac{(\lambda +3) \left(27 \lambda ^7+873 \lambda ^6+9981 \lambda ^5+27563 \lambda ^4-381408 \lambda ^3-3791260 \lambda ^2-13237104 \lambda
   -16671424\right)}{6144} \nonumber\\
&p^{(4)}_{1,3}(\lambda)=& \frac{(\lambda +3) \left(9 \lambda ^7+378 \lambda ^6+6967 \lambda ^5+72282 \lambda ^4+452252 \lambda ^3+1697424 \lambda ^2+3527616 \lambda
   +3124704\right)}{1024} \nonumber\\
&p^{(4)}_{1,4}(\lambda)=& \frac{(\lambda +3) \left(27 \lambda ^7+1017 \lambda ^6+16125 \lambda ^5+140059 \lambda ^4+722016 \lambda ^3+2215204 \lambda ^2+3751824 \lambda
   +2707264\right)}{12288} \nonumber\\
&p^{(4)}_{2,0}(\lambda)=& \frac{(\lambda +3) \left(459 \lambda ^7+19635 \lambda ^6+371775 \lambda ^5+4021241 \lambda ^4+26675766 \lambda ^3+107760380 \lambda ^2+243267960 \lambda
   +234536384\right)}{24576} \nonumber\\
&p^{(4)}_{2,1}(\lambda)=& \frac{(\lambda +3) \left(18 \lambda ^7+709 \lambda ^6+11952 \lambda ^5+110323 \lambda ^4+592866 \lambda ^3+1814348 \lambda ^2+2841112 \lambda
   +1661984\right)}{2048} \nonumber\\
&p^{(4)}_{2,2}(\lambda)=& -\frac{(\lambda +3) \left(18 \lambda ^7+769 \lambda ^6+14541 \lambda ^5+156459 \lambda ^4+1026717 \lambda ^3+4079068 \lambda ^2+9022540 \lambda
   +8515840\right)}{1024} \nonumber\\
&p^{(4)}_{2,3}(\lambda)=& -\frac{(\lambda +3) \left(54 \lambda ^7+2175 \lambda ^6+37776 \lambda ^5+365993 \lambda ^4+2131686 \lambda ^3+7453988 \lambda ^2+14474568 \lambda
   +12027488\right)}{6144} \nonumber\\
&p^{(4)}_{2,4}(\lambda)=& -\frac{(\lambda +3)^2 \left(9 \lambda ^6+302 \lambda ^5+4131 \lambda ^4+29650 \lambda ^3+118452 \lambda ^2+250968 \lambda +221120\right)}{8192} \nonumber\\
&p^{(4)}_{3,1}(\lambda)=& -\frac{(\lambda +3) \left(19 \lambda ^7+797 \lambda ^6+14783 \lambda ^5+156299 \lambda ^4+1011478 \lambda ^3+3981140 \lambda ^2+8760984 \lambda
   +8252288\right)}{1024} \nonumber\\
&p^{(4)}_{3,2}(\lambda)=& -\frac{(\lambda +3) \left(25 \lambda ^7+1031 \lambda ^6+18493 \lambda ^5+186753 \lambda ^4+1144046 \lambda ^3+4240252 \lambda ^2+8776344 \lambda
   +7795072\right)}{2048} \nonumber\\
&p^{(4)}_{3,3}(\lambda)=& -\frac{\mathit{f} (\lambda +6) \left(3 \lambda ^5+81 \lambda ^4+871 \lambda ^3+4667 \lambda ^2+12470 \lambda +13280\right)}{1024} \nonumber\\
&p^{(4)}_{3,4}(\lambda)=& -\frac{\mathit{f}^2 (\lambda +3) \left(\lambda ^3+17 \lambda ^2+90 \lambda +160\right)}{4096} \nonumber\\
&p^{(4)}_{4,0}(\lambda)=& \frac{(\lambda +3) \left(227 \lambda ^7+9363 \lambda ^6+169953 \lambda ^5+1752177 \lambda ^4+11034384 \lambda ^3+42243432 \lambda ^2+90519984 \lambda
   +83227920\right)}{49152} \nonumber\\
&p^{(4)}_{4,1}(\lambda)=& \frac{(\lambda +3) \left(38 \lambda ^7+1569 \lambda ^6+28248 \lambda ^5+287283 \lambda ^4+1778880 \lambda ^3+6687912 \lambda ^2+14078592 \lambda
   +12733680\right)}{6144} \nonumber\\
&p^{(4)}_{4,2}(\lambda)=& \frac{(\lambda +3) \left(25 \lambda ^7+1005 \lambda ^6+17427 \lambda ^5+169035 \lambda ^4+990348 \lambda ^3+3502032 \lambda ^2+6910848 \lambda
   +5858448\right)}{12288} \nonumber\\
&p^{(4)}_{4,3}(\lambda)=& \frac{\mathit{f} (\lambda +6)^3 \left(2 \lambda ^3+27 \lambda ^2+120 \lambda +177\right)}{6144} \nonumber\\
&p^{(4)}_{4,4}(\lambda)=& \frac{\mathit{f}^4}{49152} 
\end{eqnarray}
}

\section{Zero temperature limit for the 2-antisymmetric representation}
\label{app:zerotemp}

In the 2-antisymmetric representation, the zero temperature limit of the scalar propagator is given by:
\begin{equation}
\Phi=\lim_{\substack{\tau,\bar{\tau} \to \infty \\ \mbox{\small{fixed }} \mu }}
e^{\Delta\rho}\left(
\frac{1}{2}\langle1|\mathcal{O}|1\rangle \langle2|\mathcal{O}|2\rangle - 
\frac{1}{2}\langle1|\mathcal{O}|2\rangle \langle2|\mathcal{O}|1\rangle  \right),
\end{equation}
where $\Delta$ is the conformal dimension of the dual scalar operator, related to the bulk scalar mass by
$m^{2}=\Delta\,(\Delta-2)$, and $\mc O$ is the operator
\begin{equation}
\mathcal{O}=e^{e^\rho \overline{z} V^2_{-1}}\,e^{-e^\rho z V^2_{1}}\,e^{e^{2\rho} \overline{z} V^3_{2}}.
\label{zeroT}
\end{equation}
Using the fact that $V^3_{2} = V^{2}_1\star V^{2}_1$, we need the following matrix elements:
\begin{eqnarray}
&&V_{1,1}(p,q)=\langle1|   (V^2_{-1})^p  (V^2_{1})^q |1\rangle = \delta_{p,q} q ! \frac{\Gamma(q + \lambda +1)}{\Gamma(\lambda + 1)},\nonumber\\
&&V_{2,2}(p,q)=\langle2|   (V^2_{-1})^p  (V^2_{1})^q |2\rangle = \delta_{p,q} (q +1) ! \frac{\Gamma(q + \lambda +2)}{\Gamma(\lambda + 2)},\nonumber\\
&&V_{1,2}(p,q)=\langle1|   (V^2_{-1})^p  (V^2_{1})^q |2\rangle =  - \delta_{p-1,q} \frac{(q+1) !}{\sqrt{-\lambda-1}} \frac{\Gamma(q + \lambda +2)}{\Gamma(\lambda + 1)},\nonumber\\
&&V_{2,1}(p,q)=\langle1|   (V^2_{-1})^p  (V^2_{1})^q |2\rangle =   \delta_{p,q-1} \frac{(p+1) !}{\sqrt{-\lambda-1}} \frac{\Gamma(p + \lambda +2)}{\Gamma(\lambda + 1)}.
\end{eqnarray}
Plugging this expressions in the series expansion of \ref{zeroT} we have for each matrix element:
\begin{eqnarray}
\langle i | \mathcal{O} | j \rangle = \sum_{m,n,p=0}^{\infty}{ \frac{(- e^{\rho} z)^m}{m!}\frac{(\mu e^{2 \rho} \bar{z})^n}{n!}\frac{(e^{\rho}\bar{ z})^p}{p!}} V_{i,j}(p, m + 2n),
\end{eqnarray}
leading for $R_{\sm\yng(1,1)}$ in the zero temperature (and large $\rho$) limit to: 
\begin{eqnarray}
\lim_{\substack{\tau,\bar{\tau} \to \infty \\ \mbox{\small{fixed }} \mu }} R_{\sm\yng(1,1)} &=& 
\frac{1}{2}
\left[ \sum_{n=0}^{\infty}  \left(\frac{\mu \bar{z}}{z^2}\right)^n \frac{\Gamma(2 n + 1 + \lambda)}{ n! \,\Gamma(1 + \lambda)} \right]
\left[ \sum_{n=0}^{\infty}  \left(\frac{\mu \bar{z}}{z^2}\right)^n \frac{(3 + 4 n + \lambda)\Gamma(2 n + 2 + \lambda)}{ n! \, \Gamma(2 + \lambda)} \right] - \nonumber\\
&&\frac{1}{2}
\left[ \sum_{n=0}^{\infty}  \left(\frac{\mu \bar{z}}{z^2}\right)^n \frac{\Gamma(2 n + 2 + \lambda)}{ n!  \, \Gamma(2 + \lambda)} \right]
\left[ \sum_{n=0}^{\infty}  \left(\frac{\mu \bar{z}}{z^2}\right)^n \frac{(1 + 4 n + \lambda)\Gamma(2 n + 1 + \lambda)}{ n! \,\Gamma(1 + \lambda)} \right]. \nonumber\\
\end{eqnarray}
After some simple manipulation, we recover the result (\ref{eq:zt}).

\providecommand{\href}[2]{#2}\begingroup\raggedright\endgroup

\end{document}